\documentstyle[12pt]{article}
\if@twoside  m
\else
\fi
\newcommand{\ie}{{\em i.e.}}
\newcommand{\eg}{{\em e.g.}}

\newcommand{\QED}{\mbox{\rule[-1.5pt]{7pt}{11pt}}}
\newcommand{\R}{I\!\!R}
\newcommand{\OO}{{\cal O}}
\newcommand{\SS}{{\cal S}}
\newcommand{\eps}{\varepsilon}
\newcommand{\Iff}{{\em iff $\:$}}
\newcommand{\DD}{{\cal D}}
\newcommand{\WW}{{\cal W}}

\begin{document}
\title{}
\date{}
\author{}
\noindent
{\Large\bf Bound states and scattering in quantum \\ waveguides
coupled laterally through a \\ boundary window}
\vspace{5mm}
\begin{quote}
{\large P.~Exner,$^{a,b}$ P.~\v Seba,$^{a,b}$ M.~Tater,$^a$
and D.~Van\v ek$^c$}
\vspace{3mm}

{\em a) Nuclear Physics Institute, Academy of Sciences \\
CZ--25068 \v Re\v z near Prague \\ b) Doppler
Institute, Czech Technical University, \\ B\v rehov\'a 7,
CZ--11519 Prague, \\ c) Departmant of Mathematics, FNSPE, CTU, \\
Trojanova 13, CZ--12000 Prague, Czech Republic} \\
\vspace{10mm}

We consider a pair of parallel straight quantum waveguides coupled
laterally through a window of a width $\,\ell\,$ in the common boundary.
We show that such a system has at least one bound state for any
$\,\ell>0\,$. We find the corresponding eigenvalues and eigenfunctions
numerically using the mode--matching method, and discuss their
behavior in several situations. We also discuss the scattering
problem in this setup, in particular, the turbulent behavior of the
probability flow associated with resonances. The level and
phase--shift spacing statistics shows that in distinction to closed
pseudo--integrable billiards, the present system is essentially
non--chaotic. Finally, we illustrate time evolution of wave packets
in the present model.
\end{quote}
\vspace{10mm}

\section{Introduction}

Spectral and scattering properties of quantum particles whose motion
is confined to nontrivial subsets of $\,\R^n$ represented until
recently rather textbook examples or technical tools used in proofs.
There are several reasons why these problems attracted a wave of interest
in last few years . The most mathematical among them stems
from the observation that, roughly speaking, one can choose the
region in such a way that the spectrum of the corresponding Neumann
Laplacian coincides with a chosen set$^{1,2}$;
of course, the
boundary of such a region may be in general rather complicated.

On the other hand, even regions with nice boundaries may exhibit
various unexpected properties manifested, for instance, in spectra
of the corresponding Dirichlet Laplacians. A prominent example is
the existence of bound states, \ie, localized solutions to the free
Schr\"odinger equation, in infinitely stretched regions such as bent,
branched or crossed tubes of a constant cross section --- see, \eg,
Refs. 3--7;
more references are given in the review
paper$^8\!$.

\subsection{Quantum wire systems}

A strong motivation to study such bound states and related resonance
effects$^{9-11}$
comes from recent developments in semiconductor
physics, because they can be used as models of electron motion in
so--called {\em quantum wires,} \ie, tiny strips of a very pure
semiconductor material, and similar structures. Let us briefly recall
key features of such systems; more details and a guide to physical
literature can be found in Ref. 8.

Characteristic properties of the semiconductor microstructures under
consideration are {\em small size,} typically from tens to hundreds
of $\,{\rm nm}\,$, {\em high purity,} which means that the electron mean
free path can be a few $\,\mu\rm m\,$ or even larger, and {\em
crystallic structure.} In addition, boundaries of the microstrucures
consist usually of an interface between two different semiconductor
materials; the electron wavefuction are known to be suppressed there.

Behavior of an electron in such a ``mesoscopic" system structure is,
of course, governed by the many--body Schr\"odinger equation
describing its interaction with the lattice atoms including the
boundary, external fields,  and possible impurities. The mentioned
properties allow us, however, to adopt several simplifying assumptions.
As we have said the mean free path is typically two or three
orders of magnitude greater than the size of the structure; hence the
electron motion can be assumed in a reasonable approximation as
{\em ballistic,} \ie, undisturbed by impurity scattering.

The most important simplification comes from the crystallic
structure. The one--electron Hamiltonian as a Schr\"odinger operator
with a periodic potential exhibits an absolutely continuous spectrum
--- see Ref. 12,
Sec.XIII.16 --- in the solid--state physics
language one says that the electron moves in the lattice as {\em free}
with some effective mass $\,m^*$. The latter changes, of course,
along the spectrum but one can regard it as a constant when we
restrict our attention to the physically interesting part of the
valence band; recall that its value may differ substantially from the
true electron mass, for instance, one has $\,m^* = 0.067\,m_e\,$ for
GaAs which is the most common semiconductor material used in
mesoscopic devices.

This property together with the wavefunction suppression at the
interfaces makes natural to model electrons in a quantum wire system
as free (spinless) particles living in the corresponding spatial
region with the {\em Dirichlet condition} on its boundary; an
interaction term must be added only if the whole structure is placed
into an external field. This is the framework in which the mentioned
curvature--induced bound states and resonances were studied. However,
the physical conclusions one can draw from it are not restricted to
mesoscopic devices: the results are useful for description of other
``new'' quantum systems$^{13}$
and provide fresh insights into
the classical theory of electromagnetic waveguides$^{14-16}\!$.

\subsection{Motivation of the present work}

Apart from these practical reasons, the curvature--induced bound states
provide at the same time a warning example showing that an intuition
based on semiclassical concepts may fail when dealing with quantum
systems. It is well known, for instance, that low--dimensional
Schr\"odinger operators have bound states for an arbitrarily small
coupling constant as long as the potential is not repulsive in the
mean and decays sufficiently fast at infinity$^{17,18}\!$.
A common wisdom, however, is that this is rather an exception, and that
the number of bound states of a quantum system is at least {\em
roughly} proportional to the classically allowed volume of the
phase--space. The waveguide systems in question illustrate that this
is not true, because they can exhibit in principle {\em any} number
of bound states, while having {\em no} closed classical trajectories
with the exception of an obvious zero--measure set.

In this paper we are going to consider another example of that kind,
this time consisting of two straight parallel quantum waveguides.
We suppose that they have a common boundary which has a window of
a width $\,\ell\,$ allowing the particle to leak from one duct
to the other. This an idealized setup for several recently studied
quantum--wire systems --- see Refs. 19--22.
Using a variational
argument we shall show that such a system has always at least one
bound state, \ie, an isolated eigenvalue below the threshold of the
continuous spectrum. Moreover, the system can have any prescribed
number of bound states provided the window width is chosen large
enough. These conclusions follow from simple estimates; however, they
tell us nothing about the corresponding wavefunctions and more
detailed dependence of the bound--state energies on the parameters.
To this aim, we shall formulate in Section~4 a method to solve the
problem numerically using the mode--matching technique. In particular,
we shall discuss how the first eigenvalue emerges from the continuum
as the window opens.

Interesting properties of the system are not exhausted by this. The
coupling between the wavefunctions in the ``arms" and the connecting
region allows the particle to tunnel betwen different transverse modes,
so the scattering matrix is nontrivial; one naturally expects it to have
resonances with properties similar to those of the bound states below
the first transverse--mode energy. It is not easy to prove the
existence of such resonances, because there is no natural parameter
in the problem which would make it possible to tune the intermode
coupling. Neither the window width $\,\ell\,$, nor replacing the
``empty" window by an ``opaque" one with a suitable point interaction
as in Ref. 23
allow for a sensible perturbation theory, because in
both cases the unperturbed bound state disappears as the coupling is
switched out. Hence we rely again on a numerical analysis based on the
mode--matching technique; the results confirm our expectation about
the resonance character of the scattering and its dependence on
parameters of the problem.

There is one more interesting aspect. The corresponding classical
system of coupled ducts is pseudo--integrable, its phase space being
of genus three.  Other systems of that type have been recently
studied$^{24,25}\!$;
it was shown that their quantum couterparts
exhibit a chaotic behaviour.  One asks naturally whether a similar
effect can be observed here. To find the level--spacing distribution
of the bound states, a very wide window is needed to produce a large
number of eigenvalues. At the same time, the spectrum has to be
unfolded, \ie, rescaled so that the mean spacing does not change
along it. The result suggests spacing distribution is then sharply
localized around a fixed value, hence there is no chaos. This is not
surprising, since all the bound--state wavefunctions have
transversally the shape of the first mode, co effectively they
correspond to a one--dimensional system. What is less trivial is that
the spacing distribution of the scattering phase shifts also does not
witness of a fully developed chaos; this suggests that
repeated reflections are an essential ingredient of the chaotic
behavior of particles in bounded pseudo--integrable billiards.

The above mentioned scattering analysis relies on the stationary
approach. The time evolution of wave packets would deserve a separate
study. In this paper we limit ourselves to a single example: in the
concluding section we present a numerical method to solve the
corresponding time--dependent Schr\"odinger equation, which allows us
to draw some qualitative conclusions about time delay in the
scattering on the connecting window.

\section{Preliminaries}
\setcounter{equation}{0}

The system we are going to study is sketched on Fig.~1. We consider a
Schr\"odinger particle whose motion is confined to a pair of
parallel strips of widths $\,d_1,\,d_2\,$, respectively. For
definiteness we assume that they are placed to both sides of the
$\,x$--axis, and they are separated by the Dirichlet boundary
everywhere except in the interval $\,(-a,a)\;$; we shall denote
this configuration space by $\,\Omega\,$ and $\,\ell:=2a\,$.
Putting $\,\hbar^2/2m=1\,$, we may identify the particle
Hamiltonian with the Dirichlet Laplacian,
   \begin{equation} \label{Hamiltonian}
H\,\equiv\, H(d_1,d_2;\ell)\,:=\,-\Delta_D^{\Omega}\,,
   \end{equation}
on $\,L^2(\Omega)\,$ defined in the standard way --- see Ref. 12,
Sec.XIII.15 ---
since the boundary of $\,\Omega\,$ has
the segment property, it acts as the usual Laplace operator with
the Dirichlet condition at the boundary.


\begin{figure}
   \begin{picture}(120,80)
      \linethickness{2pt}
      \put(30,60){\line(1,0){350}}
      \put(30,35){\line(1,0){135}}
      \put(245,35){\line(1,0){135}}
      \put(30,20){\line(1,0){350}}
      \thinlines
      \put(205,1){\vector(0,1){70}}
      \put(165,5){\vector(1,0){80}}
      \put(245,5){\vector(-1,0){80}}
      \put(380,35){\vector(1,0){20}}
      \put(10,35){\line(1,0){20}}
      \put(35,35){\vector(0,1){25}}
      \put(35,60){\vector(0,-1){25}}
      \put(165,35){\line(0,-1){32}}
      \put(245,35){\line(0,-1){32}}
      \put(35,20){\vector(0,1){15}}
      \put(35,35){\vector(0,-1){15}}
      \put(18,45){$d_1$}
      \put(18,24){$d_2$}
      \put(210,71){y}
      \put(395,25){x}
      \put(215,7){$\ell$}
   \end{picture}

\vspace{5mm}

\caption{Laterally coupled quantum waveguides}
   \end{figure}
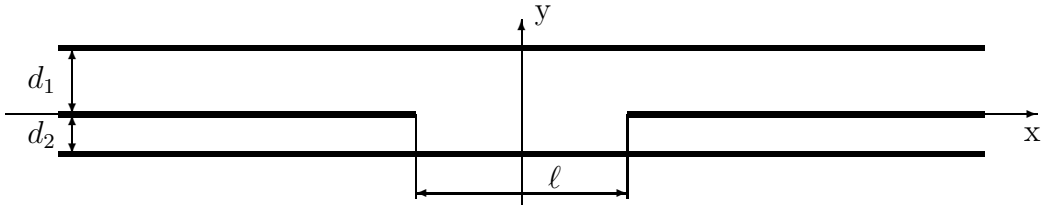


A simple bracketing argument$^{12}$
shows that
$\,H\,$ has bound states for all $\,\ell\,$ large enough. Let us
first introduce some notation. Set $\,d:=\max\{d_1,d_2\}\,$
and $\,D:=d_1+d_2\,$, and furthermore
$$
\nu\,:=\,{\min\{d_1,d_2\}\over \max\{d_1,d_2\}}\,.
$$
We shall also use $\,\mu_d:=\left(\pi\over d\right)^2\,$, and
$\,\mu_D,\,\mu_{\ell}\,$ corresponding in the same way to $\,D\,$
and $\,\ell\,$, respectively. Cutting now  $\,\Omega\,$ by
the additional Neumann or Dirichlet boundaries parallel to the
$\,y$--axis at $\,x=\pm a\,$, we get $\,H_t^{(N)}\oplus H_c^{(N)}
\le H\le H_t^{(D)}\oplus H_c^{(D)}\,$, where the ``tail'' part
corresponds to the four halfstrips and the rest to the central
part with the Neumann and Dirichlet condition on the vertical
boundaries, respectively.

Since $\,\sigma_{ess}(H_t^{(j)})= [\mu_d,\infty),\; j=N,D\,$,
the same is by the minimax principle true for $\,H\,$, and possible
isolated eigenvalues of $\,H\,$ are squeezed between those of
$\,H_c^{(j)},\; j=N,D\,$. The Neumann estimate tells us that
   \begin{equation} \label{lower bound}
\inf \sigma(H(d_1,d_2;\ell))\,\ge\,\mu_D\,=\,\mu_d
(1+\nu)^{-2}\,.
   \end{equation}
On the other hand, $\,H\,$ has an eigenvalue below
$\,\mu_D\,$ provided $\,H_c^{(D)}\,$ does, which is true if $\,\mu_D+
\mu_{\ell}\le \mu_d\;$; this shows that a sufficient condition for
$\,H(d_1,d_2;\ell)\,$ to have at least a bound state is that the
length of the opening satisfies the inequality
   \begin{equation} \label{bracketing existence}
\ell\,\ge\, {d(1+\nu) \over \sqrt{\nu(\nu+2)}}\,.
      \end{equation}
If $\,\nu=1\,$, the coefficient on the right side is
$\,2/\sqrt{3}\approx 1.155\;$; it grows as $\,\Omega\,$ becomes
more asymmetric.

More generally, the number of eigenvalues of $\,H_c^{(D)}\,$ is
$\,N_D:=\left\lbrack\sqrt{(\mu_d-\mu_D)/\mu_{\ell}}\right\rbrack\,$,
where $\,\left\lbrack\cdot\right\rbrack\,$ denotes the entire part
(recall that since $\,D\le2d\,$, the first transversally excited
state is already above $\,\mu_d\,$), while the number of the
``Neumann'' eigenvalues is $\,N_N:=1+N_D\;$; this means that the
number of bound states of $\,H(d_1,d_2;\ell)\,$ satisfies the
inequality
   \begin{equation} \label{bracketing bound}
\left\lbrack {\ell\over d}\, {\sqrt{\nu(\nu+2)}\over 1+\nu}
\right\rbrack\,\le \,N\,\le\, 1\,+\, \left\lbrack {\ell\over d}\,
{\sqrt{\nu(\nu+2)}\over 1+\nu} \right\rbrack\,.
   \end{equation}
We see that $\,H(d_1,d_2;\ell)\,$ has isolated eigenvalues, at least
for $\,\ell\,$ large enough, despite the absence of (a
nonzero--measure set of) closed classical trajectories mentioned in
the introduction. In heuristic terms, this may be understood as a
manifestation of the fact that the semi-infinite ``spikes'' of the
opened barrier between the two ducts are capable of reflecting a {\em
quantum} particle due to a finite smearing of the wavepacket. In the
same way, one finds that the $\,m$th eigenvalue $\,\mu_m\,$ of
$\,H(d_1,d_2;\ell)\,$ is estimated by
   \begin{equation} \label{eigenvalue bound}
\left({m-1\over \lambda}\right)^2 \,\le\, {\mu_m\over\mu_d}\,-\,
{1\over (1+\nu)^2}\,\le\, \left({m\over \lambda}\right)^2\,,
   \end{equation}
where $\,\lambda:=\ell/d\,$, and that the critical value
$\,\lambda_m\equiv\ell_m/d\,$ at which $\,m$th eigenvalue appears
satisfies the bounds
   \begin{equation} \label{critical bound}
{(m-1)(1+\nu) \over \sqrt{\nu(\nu+2)}}\,\le\, \lambda_m\,\le\,
{m(1+\nu) \over \sqrt{\nu(\nu+2)}}\,.
   \end{equation}
To learn more about the dependence of the eigenvalues and the
corresponding eigenfunctions on $\,\lambda\,$ and $\,\nu\,$, we have
to use a different technique.

\section{Existence of bound states}
\setcounter{equation}{0}

The above existence argument giving (\ref{bracketing existence}) is
a crude one; in fact, there is no lower bound on the window width
as the following result shows:
\vspace{3mm}

\noindent
{\bf Theorem:} \quad $\, H(d_1,d_2;\ell)\,$ has an isolated
eigenvalue in $\,[\mu_D,\mu_d)\,$ for any $\,\ell>0\,$.
\vspace{3mm}

\noindent
{\em Proof:} \quad We modify for the present purpose the
variational argument of Ref.~6;
see also Ref.~8, Sec.2.
Without loss of generality we may assume that $\,d_2\le d_1=d\,$.
The transverse ground--state wavefunction is then
$$
\chi_1(y)\,:=\, \left\lbrace\: \begin{array}{lll}
\sqrt{2\over d_1}\,\sin(\kappa_1 y) & \dots & y\in(0,d_1) \\ \\
0 & \dots & {\rm otherwise}    \end{array} \right.
$$
where $\,\kappa_1:= \sqrt{\mu_d}\;$; similarly we define the
transverse ground state in the opening,
$$
\eta_1(y)\,:=\,\sqrt{2\over D}\,\sin(K_1(d_1-y))
$$
with $\,K_1:= \sqrt{\mu_D}\,$. For any $\,\Phi\in D(H)\,$
we put
$$
q[\Phi]\,:=\, \|H\Phi\|^2- \mu_d\|\Phi\|^2
$$
(if not marked explicitly, the norms always refer to
$\,L^2(\Omega)\,$).

Since the essential spectrum of $\,H\,$ starts at $\,\mu_d\,$,
we have to find a trial function $\,\Phi\,$ such that
$\,q[\Phi]<0\,$; it has to belong to the form domain
$\,Q(H)\,$ which means, in particular, that it must be continuous
inside $\,\Omega\,$ but not necessarily smooth. Notice first that if
$\,\Phi(x,y)= \varphi(x) \chi_1(y)\,$, we have
   \begin{equation} \label{q form}
q[\Phi]\,=\, \|\varphi'\|^2_{L^2(\R)}\,.
   \end{equation}
To make the longitudinal contribution to the kinetic energy
small, we use an external scaling. We choose an interval
$\,J:=[-b,b]\,$ for a positive $\,b>a\,$ and a function
$\,\varphi\in\SS(\R)\,$ such that $\,\varphi(x)=1\,$ if
$\,x\in J\;$; then we define the family $\,\{ \varphi_{\sigma}
:\: \sigma>0\,\}\,$ by
$$
\varphi_{\sigma}(x)\,:=\, \left\lbrace\: \begin{array}{lll}
\varphi(x) & \dots & |x|\le b \\ \\
\varphi(\pm b+\sigma(x\mp b)) & \dots & |x|\ge b    \end{array}
\right.
$$
Finally, let us choose a localization function $\,j\in
C_0^{\infty}((-a,a))\,$ and define
   \begin{equation} \label{trial}
\Phi_{\sigma,\eps}(x,y)\,:=\, \varphi_{\sigma}(x)\,
[\chi_1(y)+\eps j(x)^2 \eta_1(y)]
   \end{equation}
for any $\,\sigma,\,\eps>0\,$.
The main point of the construction is that we modify the
factorized function we started with in two mutually disjoint
regions, outside and inside the rectangle $\,J\times
(-d_2,d_1)\,$. Hence the functions $\,\varphi'_{\sigma}\,$
and $\,j^2\,$ have disjoint supports. Using this together
with the identity
$$
\|\varphi'_{\sigma}\|^2_{L^2(\R)}\,=
\sigma\|\varphi'\|^2_{L^2(\R)}
$$
and the explicit forms of the functions $\,\chi_1,\,\eta_1\,$,
we substitute (\ref{trial}) into (\ref{q form}) and
find after a tedious but straightforward computation
   \begin{eqnarray} \label{trial value}
\lefteqn{q[\Phi_{\sigma,\eps}]\,=\,
\sigma\|\varphi'\|^2_{L^2(\R)}\,-\, 4\pi\eps d_1^{-3/2}
D^{-1/2} \|j^2\|^2_{L^2(\R)}\, \sin\left( \pi\over 1+\nu
\right) } \nonumber \\ \\ && \phantom{MMM}
+\, \eps^2\left\lbrace\, \|2jj'\|^2_{L^2(\R)}
-(\mu_d^2-\mu_D^2)\, \|j^2\|^2_{L^2(\R)}\, \right\rbrace\,.
\nonumber
   \end{eqnarray}
By construction, the last two terms on the right side of
(\ref{trial value}) are independent of $\,\sigma\,$.
Moreover, the term linear in $\,\eps\,$ is negative
(recall that $\,\nu\in(0,1]\,$), so choosing $\,\eps\,$
sufficiently small, we can make it dominate over the
quadratic one. Finally, we fix this $\,\eps\,$ and
choose a small enough $\,\sigma\,$ to make the right
side of (\ref{trial value}) negative. \quad \QED
\vspace{3mm}

\noindent
{\bf Remark:} \quad Though it is not the subject of the
present paper, we want to note that the same argument
demonstrates existence of a bound state in a straight
Dirichlet strip with an arbitrarily small protrusion;
one has only to replace $\,J\times [-d_2,0]\,$ by a
rectangle contained in the protruded part. An alternative proof of
this result has been given recently in Ref.~26;
these autors also
derived an asymptotic formula for the eigenvalue in terms of the
protrusion volume.

\section{Mode matching}
\setcounter{equation}{0}

To learn more about the eigenvalues and eigenfunctions in
question, we shall now solve the corresponding Schr\"odinger
equation numerically. Since $\,\Omega\,$ consists of several
rectangular regions, the easiest way to do that is by the
mode--matching method.

\subsection{The symmetric case}

Consider first the situation when $\,d_1=d_2=d\,$. The Hamiltonian
(\ref{Hamiltonian}) then decouples into an orthogonal sum of the
even and the odd part, the spectrum of the latter being clearly
trivial, \ie, the same as in the case $\,\ell=0\,$. At the same
time, the mirror symmetry with respect to the $\,y$--axis allows
us to consider separately the symmetric and antisymmetric
solutions.

We may therefore restrict ourselves to the part of
$\,\Omega\,$ in the first quadrant, with the Neumann boundary
condition in the segment $\,(0,a)\,$ of the $\,x$--axis, and
Neumann or Dirichlet condition in the segment $\,(0,d)\,$ of
the $\,y$--axis. We expand the sought solutions in terms of
corresponding transverse eigenfunctions
   \begin{eqnarray} \label{transverse eigenfunctions}
\chi_j(y) &\!:=\!& \sqrt{2\over d}\,\sin(\kappa_j y)\;,\quad
j=1,2,\dots \\ \nonumber \\
\phi_j(y) &\!:=\!& \sqrt{2}\,\eta_{2j-1}(y) \,=\,
\sqrt{2\over d}\,\sin(K_{2j-1}(d-y))\;,\quad j=1,2,\dots\,,
   \end{eqnarray}
where $\,\kappa_j:=j\kappa_1\,$ and $\,K_{2j-1}:= (2j-1)K_1\,$.
A natural Ansatz for the solution of an energy $\,\epsilon\mu_d\,,
\; {1\over 4}\le \epsilon<1\,$, is
   \begin{equation} \label{Ansatz tube}
\psi(x,y)\,=\, \sum_{j=1}^{\infty}\, b_j\, e^{q_j(a-x)}\, \chi_j(y)
   \end{equation}
for $\,x\ge a\,$, where $\,q_j:=\kappa_1\sqrt{j^2-\epsilon}\,$, and
   \begin{equation} \label{Ansatz window}
\psi_s(x,y)\,=\, \sum_{j=1}^{\infty}\, a_j\,{\cosh(p_jx)
\over \cosh(p_ja)}\, \phi_j(y)\;,\quad
\psi_{as}(x,y)\,=\, \sum_{j=1}^{\infty}\, a_j\,{\sinh(p_jx)
\over \sinh(p_ja)}\, \phi_j(y)
   \end{equation}
for $\,0\le x\le a\,$ and the symmetric and antisymmetric cases,
respectively, where the longitudinal momentum is defined by
$\,p_j:=\kappa_1\sqrt{\left(j-{1\over 2}\right)^2-\epsilon}\,$.
It is straightforward to compute the norms of the functions
(\ref{Ansatz tube}) and (\ref{Ansatz window}); since
$\,j^{-1}q_j\,$ and $\,j^{-1}p_j\,$ tend to $\,\mu_d\,$ as
$\,j\to\infty\,$, the square integrability of $\,\psi\,$
requires the sequences $\,\{a_j\}\,$ and $\,\{b_j\}\,$ to
belong to the space $\,\ell^2(j^{-1})\,$.

As an element of the domain of $\,H$, the function $\,\psi\,$
should be continuous together with its normal derivative at
the segment dividing the two regions, $\,x=a\,$. Let us first
solve this condition formally. The continuity means
$\,\sum_{k=1}^{\infty} a_k\phi_k(y)= \sum_{k=1}^{\infty}
b_k\chi_k(y)\;$; using the orthonormality of $\,\{\chi_j\}\,$
we get from here
   \begin{equation} \label{continuity}
b_j\,=\, \sum_{k=1}^{\infty}\, a_k (\chi_j,\phi_k)\,.
   \end{equation}
In the same way, the normal--derivative continuity at $\,x=a\,$
yields
   \begin{equation} \label{derivative continuity}
q_jb_j\,+\, \sum_{k=1}^{\infty}\, a_k p_k \tanh(p_ka)\,
(\chi_j,\phi_k)\,=\,0
   \end{equation}
in the Neumann case, and the analogous relation with $\,\tanh\,$
replaced by $\,\coth\,$ for Dirichlet. Substituting from
(\ref{continuity}) to (\ref{derivative continuity}), we can
write the equation as
   \begin{equation} \label{spectral C}
Ca\,=\,0\,,
   \end{equation}
where
   \begin{equation} \label{C}
C_{jk}\,:=\,\left( q_j+p_k \left\{ \begin{array}{c}
\tanh \\ \coth    \end{array} \right\} (p_ka)\, \right)\,
(\chi_j,\phi_k)
   \end{equation}
in the Neumann and Dirichlet case, respectively, with the
two orthonormal bases related by
\begin{equation} \label{overlap integral}
(\chi_j,\phi_k)\,=\, {(-1)^{j-k}\over\pi}\, {2j\over {j^2-
\left(k-{1\over 2}\right)^2}}\,.
   \end{equation}

One has to make sure, of course, that the equation (\ref{C})
makes sense, and that one can solve it by a sequence of
truncations. It is possible to follow the procedure formulated
in Ref.~4.
A more direct way, however, is to notice that
if $\,\psi\,$ is an eigenvector of $\,H\,$, it must belong to
the domain of any integer power of this operator. It is easy
to check that $\,\psi\in D(H^n)\;$ \Iff $\;\{a_j\},\,\{b_j\}
\in \ell^2(j^{2n-1})\;$; hence the sought sequences should
belong to $\,\ell^2(j^s)\,$ for all $\,s\ge -1\,$. This fact
also justifies {\em a posteriori} the interchange of summation and
differentiation we have made in the matching procedure.

Consider now the diagonal operator $\,S_r\,$ on $\,\ell^2(j^{-1})\,$,
$\,(S_r a)_j:= j^{-r}a_j\,$. If $\,C\,$ has zero eigenvalue with
a fast decaying eigenvector, the same is true for $\,C^{(s,r)}:=
S_s CS_r\,$ with arbitrary non--negative $\,s,r\,$. The last
named operator is represented by the matrix
$$
C^{(s,r)}_{jk}\,:=\, \left( q_j+p_k \tanh(p_ka)\, \right)\,
{(-1)^{j-k}\over\pi}\, {2j^{1-s} k^{-r}\over {j^2-
\left(k-{1\over 2}\right)^2}}
$$
(for the sake of brevity, we speak about the Neumann case only),
so it is Hilbert--Schmidt for $\,r,s\,$ large enough, and its
eigenvalues can therefore be obtained from a sequence of truncated
operators. Since finite matrices pose no convergence problems, the
truncation procedure may be applied to the operator $\,C\,$ directly.

Of course, $\,C^{(r,s)}\,$ may have eigenvectors to which no
square--summable eigenvector of $\,C\,$ corresponds, because
$\,S_r^{-1}\,$ is unbounded for $\,r>0\,$. Fortunately, the search
for solutions may be terminated once we find the number of them which
saturates the upper bound derived in Section~2.

\subsection{An alternative method}

A natural modification of the above described procedure is to
express $\,\{a_k\}\,$ from (\ref{continuity}) using the
orthonormality of $\,\{\phi_k\}\,$, and to substitute it into
(\ref{derivative continuity}); then the spectral condition
acquires the form
   \begin{equation} \label{spectral K}
b+Kb\,=\,0\,,
   \end{equation}
where
   \begin{equation} \label{K}
K_{jm}\,:=\, {1\over q_j}\: \sum_{k=1}^{\infty}\,
(\chi_j,\phi_k)\,p_k \tanh(p_ka)\, (\phi_k,\chi_m)\,,
   \end{equation}
and the same with $\,\coth(p_ka)\,$ in the Dirichlet case.

The two approaches are, of course, equivalent. Solving the
equation numerically, however, we truncate not only the
matrices but also the series in (\ref{K}). The sequences
approximating a given eigenvalue are therefore different.
Moreover, in the examples given below we find them
monotonous in the opposite sense. The sequences coming
from (\ref{spectral C}) were approaching the limiting
values from above, while those obtained from
(\ref{spectral K}) were increasing; in combination this
gives a good idea about the numerical stability of the
solution.

\subsection{The asymmetric case}

Let us pass now to the case, when the widths of the ducts
are nonequal, $\,d_1\ne d_2\,$. Without loss of generality,
we may again suppose that $\,d_2\le d_1=d\,$. With the
mirror symmetry with respect to the $\,y$--axis in mind,
we shall consider the right--halfplane part of $\,\Omega\,$
only with the Neumann and Dirichlet condition on the
segment $\,\WW:=[-d_2,d_1]\,$ of the $\,y$--axis.

To expand the sought solution, we need again suitable
transverse bases. In the ``connecting part'', $\,0\le
x\le a\,$, we use
   \begin{equation} \label{window basis}
\eta_k(y) \,=\,
\sqrt{2\over D}\,\sin(K_k(d_1-y))\;,\quad k=1,2,\dots\,,
   \end{equation}
where $\,K_k:=kK_1= k\kappa_1(1+\nu)^{-1}\,$. On the other
hand, for the ducts we choose
   \begin{eqnarray} \label{duct bases}
\chi_j^{(+)}(y) &\!:=\!& \sqrt{2\over d_1}\,
\sin(\kappa_j y)\, i_{+}(y)\;,\quad
j=1,2,\dots\,, \\ \nonumber \\
\chi_j^{(-)}(y) &\!:=\!& -\,\sqrt{2\over d_2}\,
\sin(\kappa_j\nu^{-1}y)\, i_{-}(y)\;,\quad
j=1,2,\dots\,,
   \end{eqnarray}
where $\,\kappa_j:=j\kappa_1\,$ and $\,i_{\pm}\,$ are
the indicator functions of the intervals $\,\DD_{+}:=
[0,d_1]\,$ and $\,\DD_{-}:= [-d_2,0]\,$, respectively.

The union of the two bases is, of course, an orthonormal
basis in $\,L^2(\WW)\,$. Since the numerical computation
involves a truncation procedure, we need to introduce
a proper ordering. For that we arrange the eigenvalues
corresponding to (\ref{duct bases}) to a single
nondecreasing sequence. Equivalently, we arrange the
numbers $\,j,\,k\nu^{-1}$ with $\,j,k=1,2,\dots\,$ into a
nondecreasing sequence (if $\,\nu\,$ is rational and
there is a coincidence, any order can be chosen in the
pair); we denote its elements by $\,\theta_m\,$,
$$
\theta_1:=1\,,\quad \theta_2:=\min\{2,\nu^{-1}\}\,,
\quad {\rm etc.}
$$
The corresponding ordered basis in $\,L^2(\WW)\,$ is
   \begin{equation} \label{common duct basis}
\xi_m\,:\;\xi_m(y) \,=\, \, \left\lbrace\: \begin{array}{lll}
\chi_j^{(+)}(y) & \dots & \theta_m=j \\ \\
\chi_j^{(-)}(y) & \dots & \theta_m=j\nu^{-1}
   \end{array} \right.
   \end{equation}
Consider first the even solutions, \ie, the Neumann condition
at $\,x=0\,$. A natural Ansatz for a solution of an energy
$\,\epsilon\mu_d\,,\; (1+\nu^{})^{-2}\le \epsilon<1\,$, is
   \begin{eqnarray} \label{Neumann Ansatz}
\psi(x,y)&\!:=\!& \sum_{k=1}^{\infty}\, a_k\,{\cosh(p_kx)
\over \cosh(p_ka)}\, \eta_k(y)\;\qquad \dots \quad
0\le x\le a\,, \nonumber \\ \\
\psi(x,y)&\!:=\!& \sum_{j=1}^{\infty}\, b_j^{(\pm)}\,
e^{q_j^{(\pm)}(a-x)}\, \chi_j^{(\pm)}(y)\;\quad
\dots \quad x\ge a\,;\; y\in\DD_{\pm}\,, \nonumber
   \end{eqnarray}
where
$$
p_j\,:=\,\kappa_1\,\sqrt{\left(j\over{1+\nu}\right)^2
-\epsilon}\,,
$$
and
$$
q_j^{(+)}\,:=\,\kappa_1\,\sqrt{j^2-\epsilon}\;,\quad
q_j^{(-)}\,:=\,\kappa_1\,\sqrt{\left(j\over{\nu}\right)^2
-\epsilon}\,.
$$
The duct part of (\ref{Neumann Ansatz}) can be also written
in a unified way as
   \begin{equation} \label{duct Ansatz}
\psi(x,y)\,=\, \sum_{m=1}^{\infty}\, c_m\,
e^{r_m(a-x)}\, \xi_m(y)\,,
   \end{equation}
where
$$
c_m\,:=\, \, \left\lbrace\: \begin{array}{lll}
b_j^{(+)} & \dots & \theta_m=j \\ \\
b_j^{(-)} & \dots & \theta_m=j\nu^{-1}
   \end{array} \right. \qquad {\rm and} \qquad
r_m\,:=\, \, \left\lbrace\: \begin{array}{lll}
q_j^{(+)} & \dots & \theta_m=j \\ \\
q_j^{(-)} & \dots & \theta_m=j\nu^{-1}
   \end{array} \right.
$$
Using the continuity of the function and its normal derivative
at $\,x=a\,$ together with the orthonormality of $\,\{\chi_j
^{(\pm)}\}\,$, we find conditions for the coefficient
sequences,
   \begin{eqnarray} \label{as matching}
b_j^{(\pm)} &\!\!=\!\!& \sum_{k=1}^{\infty}\, a_k\,
(\chi_j^{(\pm)},\eta_k)\,, \\ \nonumber \\
q_j^{(\pm)}b_j^{(\pm)} &\!\!+\!\!& \sum_{k=1}^{\infty}\,
a_k\,p_k\,\tanh(p_ka)\,(\chi_j^{(\pm)},\eta_k)\,=\,0\,.
   \end{eqnarray}
This can be also written as
$$
c_m\,=\, \sum_{k=1}^{\infty}\, a_k\,(\xi_m,\eta_k)\;,
\quad r_mc_m\,+\, \sum_{k=1}^{\infty}\,
a_k\,p_k\,\tanh(p_ka)\,(\xi_m,\eta_k)\,=\,0\;;
$$
substituting from the first equation to the second one,
we obtain the spectral condition in the form (\ref
{spectral C}) with
   \begin{equation} \label{as C}
C_{mk}\,:=\, (r_m +p_k\tanh(p_ka))\, (\xi_m,\eta_k)\,,
   \end{equation}
where the overlap integrals are given by
\begin{eqnarray} \label{as overlap integral}
(\chi_j^{(+)},\eta_k) &\!=\!& {2j\over\pi\sqrt{1+\nu}}\:
{\sin\left(\pi k\over 1+\nu \right)\over {j^2-
\left(k\over{1+\nu}\right)^2}}\,, \nonumber \\ \\
(\chi_j^{(-)},\eta_k) &\!=\!& {2j\over\pi}\,
\sqrt{\nu\over{1+\nu}}\:
{\sin\left(\pi k\over 1+\nu \right)\over {j^2-
\left(k\nu\over{1+\nu}\right)^2}}\,. \nonumber
   \end{eqnarray}
In the odd case, \ie, Dirichlet condition at $\,x=0\,$,
we get the same equation with $\,\tanh\,$ replaced by
$\,\coth\,$ in (\ref{as C}).

By a straightforward modification of the above argument,
one can check that the coefficient sequences have a
faster--than--powerlike decay and the spectral condition
can be solved by a sequence of truncations. One can also
rewrite the condition in the form analogous to
(\ref{spectral K}), $\,c+Kc=0\,$, where
   \begin{equation} \label{as K}
K_{jm}\,:=\, {1\over r_j}\: \sum_{k=1}^{\infty}\,
(\xi_j,\eta_k)\,p_k \tanh(p_ka)\, (\eta_k,\xi_m)\,.
   \end{equation}

\section{Scattering}
\setcounter{equation}{0}

The analysis is similar to that of the previous section. The incident
wave is supposed to be of the form $\,\chi_j^{(+)}(y)\,
e^{-ik_j^{(+)}x}\,$ in the upper channel, where we have introduced
$$
k_j^{(+)}\,:=\,\kappa_1\,\sqrt{k^2\!-j^2}\;,\quad
k_j^{(-)}\,:=\,\kappa_1\,\sqrt{k^2\!-\left(j\over{\nu}\right)^2}\;;
$$
we denote by $\,r_{jj'}^{(\pm)},\, t_{jj'}^{(\pm)}\,$, respectively, the
corresponding reflection and transmission amplitudes to the $\,j'$--th
transverse mode in the upper/lower guide. Due to the mirror symmetry,
we can again separate the symmetric and antisymmetric situation with
respect $\,x=0\,$ and to write
   \begin{equation} \label{rt coefficients}
r_{jj'}^{(\pm)}=\, {1\over 2}\, \left(\rho_{jj'}^{(s,\pm)}+
\rho_{jj'}^{(a,\pm)}\right)\,, \quad
t_{jj'}^{(\pm)}=\, {1\over 2}\, \left(\rho_{jj'}^{(s,\pm)}-
\rho_{jj'}^{(a,\pm)}\right)\,,
   \end{equation}
where $\,\rho_{jj'}^{(\sigma,\pm)}\,,\; \sigma=s,a\,$, are the
appropriate reflection amplitudes. In the symmetric case we have
the following Ansatz for the solution
   \begin{eqnarray} \label{Neumann scattering Ansatz}
\psi(x,y)&\!:=\!& \sum_{\ell=1}^{\infty}\, a_{\ell}\,{\cos(p_{\ell}x)
\over \cos(p_{\ell}a)}\, \eta_{\ell}(y)\;\hspace{48mm} \dots
\quad 0\le x\le a\,, \nonumber \\ \nonumber \\
\psi(x,y)&\!:=\!& \sum_{j'=1}^{\infty} \left(\,
\delta_{jj'}\, e^{-ik_j^{(+)}(x-a)}\!+\! \rho_{jj'}^{(+)}
e^{ik_{j'}^{(+)}(x-a)}\,\right)\, \chi_{j'}^{(+)}(y)
\quad \dots \quad x\ge a\,;\, y\in\DD_+\,, \nonumber \\ \\
\psi(x,y)&\!:=\!& \sum_{j'=1}^{\infty}\, \rho_{jj'}^{(-)}
e^{ik_{j'}^{(-)}(x-a)}\, \chi_{j'}^{(-)}(y)
\hspace{40mm} \dots \quad
x\ge a\,;\; y\in\DD_-\,. \nonumber
   \end{eqnarray}
The last two relations can be written also as
$$
\psi(x,y)\,=\, \sum_{m'=1}^{\infty} \left(\,
\delta_{mm'}\, e^{-ik_m(x-a)}\!+\! \rho_{mm'}
e^{ik_{m'}(x-a)}\,\right)\, \xi_{m'}(y) \,,
$$
where
$$
\rho_{mm'}\,:=\, \, \left\lbrace\: \begin{array}{lll}
\rho_{jj'}^{(+)} & \dots & \theta_m=j\,,\, \theta_{m'}=j' \\ \\
\rho_{jj'}^{(- )} & \dots & \theta_m=j\,,\, \theta_{m'}=j'\nu^{-1}
   \end{array} \right. \qquad
k_m\,:=\, \, \left\lbrace\: \begin{array}{lll}
k_j^{(+)} & \dots & \theta_m=j \\ \\
k_j^{(-)} & \dots & \theta_m=j\nu^{-1}
   \end{array} \right.
$$
Matching the functions (\ref{Neumann scattering Ansatz}) smoothly at
$\,x=a\,$ we arrive in the same way as above at the equation
   \begin{equation} \label{symmetric scattering}
\sum_{m'=1}^{\infty} \left(\, ik_\ell+ p_{m'}\tan(p_{m'}a)
\,\right)\, (\xi_\ell,\eta_{m'})\,a_{m'}\,=\, 2ik_\ell\delta_{m\ell}\,,
   \end{equation}
where the index $\,m\,$ corresponds to the incident wave and the overlap
integrals are given again by (\ref{as overlap integral}); in the
antisymmetric case on has to replace $\,\tan\,$ by $\,-\cot\,$. The
reflection amplitudes are given then by
   \begin{equation} \label{symmetric reflection amplitudes}
\rho_{m\ell}^{(\pm)}\,=\,-\delta_{m\ell}\,+\, \sum_{m'=1}^{\infty}
a_{m'}^{(\pm)}\,(\xi_\ell,\eta_{m'})\;;
   \end{equation}
they determine the full S--matrix via (\ref{rt coefficients}).

\section{The results}
\setcounter{equation}{0}

\subsection{Bound states}

The results of the mode--matching computation are illustrated on
Figures 2--4.  In accordance with the general results of Section~2
the eigenvalues decrease mono\-tonous\-ly with the increasing window
width and one can sandwich them between the estimates
(\ref{eigenvalue bound}). The eigenfunctions decay exponentially out
of the ``interaction" region.  The ground state wavefunction is, of
course, positive up to a phase factor; the nodal lines of the excited
states are parallel to the $\,y$--axis. The last feature illustrates
once more that apart of the exponential tails in the ducts, the
quantum particle ``feels" the window part as a closed rectangular
resonator.

It is also interesting to estimate the rate at which the eigenvalues
emerge from the continuum. The results of the mentioned paper$^{26}$
together with the Dirichlet bracketing allow us to find a simple
upper bound for the ground--state energy by means of a single strip
with a ``blister" whose volume is squeezed to zero. Since the
asymptotic formula derived in Ref.~26 applies to ``gentle"
protrusions, it may be employed if the power with which the bump is
scaled transversally is larger than the longitudinal one. Hence the
gap between the eigenvalue and the continuum for a narrow window is
bound from below by $\,C(\eps)\ell^{4+\eps}+ \OO(\ell^5)\,$ for any
$\,\eps>0\,$.

This can be compared with the numerical results. Redrawing the first
eigenvalue curve of Fig.~2 and analogous results for $\,\nu\neq 1\,$
in the logarithmic scale, we find that the asymptotic behavior is
powerlike. The convergence of our method for small $\,\ell\,$ is
rather slow; nevertheless, using cut--off dimensions of order
$\,10^3\,$ we get for the power values witnessing clearly that the
above bound is saturated,
   \begin{equation} \label{narrow window asymptotics}
\mu_1(\ell)\,=\,\mu_d-\,c(\nu)\ell^4+\,\OO(\ell^5)\,.
   \end{equation}
The numerically found coefficient $\,c(\nu)\,$ is monotonous and
reaches its maximum value for $\,\nu=1\;$; this is the expected
behavior as can be seen from a simple bracketing argument.  Proving
the conjecture (\ref{narrow window asymptotics}) and finding an
analytical expression for $\,c(\nu)\,$ remains an open problem; the
same can be said about the ``coupling--constant thresholds'', \ie,
the way the other eigenvalues emerge from the continuum.

\subsection{Scattering}

The passage of the particle through the window region is determined
by the transmission and reflection amplitudes (\ref{rt
coefficients}). The physically interesting quantity is the
conductivity. If we suppose, for instance, that the particle comes
from the upper right guide and leaves through the upper left one,
then the conductivity (denoted conventionally as TP and measured in
the standard units $\,2\,e^2/h\,$) is given by
   \begin{equation} \label{conductivity}
G(k)\,=\, \sum_{j,j'=1}^{[k]}\, {k_{j'}^{(+)}\over k_j^{(+)}}\:
|t_{jj'}^{(+)}(k)|^2\,,
   \end{equation}
and similarly for the other combinations; the summation runs over all
open channels. The resonance structure is visible on Figure~5.

Another insight can be obtained by investigating the probability flow
distribution associated with the generalized eigenvector
(\ref{Neumann scattering Ansatz}) which is defined in the standard way,
   \begin{equation} \label{probability flow}
\vec j(\vec x)\,:=\, -i\bar\psi(\vec x) \vec\nabla\psi(\vec x)\,.
   \end{equation}
The flow patterns change with the
momentum of the incident particle. They exhibit conspicuous vortices
at the resonance energies which represent the ``trapped part'' of the
wavefunction; this phenomenon is illustrated on Figure~7. It has been
argued in the literature that leaky wires similar to those studied
here may serve as switching devices$^{22}\!$.
The vortices
which emerge in resonance situations lead to the appearance of a
magnetic dipole moment, which might be in principle measured
experimentally. In this respect situations with a single well
developed vortex such as the one illustrated on Figure~6 are
particularly promising.

\subsection{Chaos}

Discussing a chaotic behavior of a quantum system, it is useful
to start with its classical counterpart, and in particular, its phase
space. In the present case of an infinite two--strip ``billiard''
there are no closed classical trajectories with exception of the
obvious zero--measure set, hence one has to consider the
scattering, \ie, motion of a point particle bouncing its way
through the system; the reflection from the walls is supposed be
perfectly elastic.

There are two integrals of motion: the longitudinal component
of the momentum, $\,I_1=p_x\,$, and the {\em modulus} of its
transverse part, $\,I_2=|p_y|\,$. Hence the phase space trajectory of
the system is restricted to a two--dimensional manifold (invariant
surface) in the four--dimensional phase space. However, due to the
singularity of corresponding classical flow at the edges of the
connecting window, the topology of this surface is not equivalent to
that of a two--dimensional cylinder, but rather of a pair of mutually
crossed cylinders; similar systems are usually dubbed
pseudo--integrable$^{27}\!$.
The topological structure of the invariant
surface has a consequence for the quantum counterpart: the system
cannot be quantized semiclassically.

On the other hand, the quantum system of coupled waveguides has in
view of our previous arguments bound states, even many of them \Iff
$\,\ell\sqrt{\nu} \gg d\,$. Then one can plot the distribution of the
eigenvalue spacing as shown on Figure~8 for a particular value of
$\,\nu\,$; the character of the distribution does not change as the
latter is varied. The natural unfolding means in this case to employ
the corresponding momentum value $\,ip_1=
\sqrt{\mu_d\left(\epsilon-(1\!+\!\nu)^{-2} \right)}\,$. The results
differ from typical (unfolded) eigenvalue distributions in billiards,
both integrable and chaotic, in the first place due to the existence
of the sharp localization around a fixed value. The used statistics
(several thousand eigenvalues) does not allow us to tell what is the
behavior around zero; we see, however, that the decay off the peak is
at least exponential. This differs substantially from a typical
behavior of chaotic systems, however, one should not be surprised
because all the corresponding eigenfunctions are dominated
transversally by the lowest mode, so the bound--state family in our
``billiard'' is effectively one--dimensional.

It is less trivial whether a chaotic behavior may be manifested in the
scattering; recall that spatially restricted pseudo--integrable
billiards are known to exhibit the so--called wave chaos$^{24}\!$.
To decide whether a quantum scattering system is chaotic or not, one
has to study eigenvalue distribution of the corresponding S--matrix,
again properly unfolded, which is expected to conform with that of
the Dyson circular ensemble of random matrices$^{28}$
in the former case. We
have performed this task for the system under consideration
numerically, analyzing the distribution of the spacing between two
neighboring eigenvalues of the S-matrix. The result is plotted on
Figure~9; they are compared with the Wigner and the Poissonian
distributions peculiar for the chaotic and non--chaotic situation,
respectively. It can be seen that the overall shape of this
distribution matches the Poissonian distribution for all spacings
large enough; on the other hand, the deformation of the distribution
near the origin provides a clear sign of non--integrability of the
system. The fact that this non--integrability differs from a typical
chaotic behavior can be attributed to the fact that the scattered
particle passes the window region ``only once" without being bounced to
and fro as it is the case of finite billiards.

The absence of the fully developed chaos in the coupled waveguides
can also be seen when plotting the coefficients $\,a_\ell\,$ which
determine the wavefunction in the interaction region by (\ref{Neumann
scattering Ansatz}) as illustrated on Figure~10. Their distribution
remains well localized even for higher energies of the incoming
particle, its tail being approximately exponential, while in case of
an irregular scattering one would expect a slower decay.

\section{Time evolution}

Up to now we have discussed the coupled waveguide system from the
stationary point of view only. Let us look briefly how the window
coupling can affect propagation of wavepackets in the ducts. This
problem has a natural motivation: it has been suggested recently
$^{22,29,30}$
that coupled electron waveguides provide an analogue
of the optical directional coupler in the sense that they may switch
electrons from one quantum wire to another. Moreover, the authors of
Ref.~22
conjectured that the electron switching process should
be rather fast due to the direct character of the corresponding
resonance, since the electron is not trapped in the interaction
region during the resonant switching.

The existence of probability--flow vortices discussed above in the
interaction region indicates that this might not be the case, \ie,
that the electron dwelling time in the junction may not be generally
neglected. To get a better insight we have investigated time
evolution of wave packets numerically. This can be achieved by
approximating the evolution operator by a Trotter--formula product
--- see Ref.~12, Sec.VIII.8 ---
with the Dirichlet boundary condition replaced
by a very steep and narrow potential barrier localized along the
boundary; the latter has been chosen in such a way that the dynamics
of the system was equivalent to the dynamics of the true Dirichlet
problem for all times taken into account, \ie, that the tunneling
leak was negligible during that period.

The kinetic-- and potential--part factors of the evolution operator are
then multiplication operators in the momentum and coordinate
representation, respectively; the passage between the two
representations has been realized by means of the two--dimensional
Fast Fourier Transform method$^{31}$
with a grid of $2^9\times
2^7$ points. The time evolution of a wavepacket approaching the
junction through the upper right arm of the structure is plotted on
Figure~11. The incoming wavefunction was chosen as
$\,\psi(x,y):= g(x)\chi_1^{(+)}(y)\,$, where $\,g(x):=
\exp\{-a(x\!-\!x_0)^2\!+i k x\}\,$ with suitably chosen parameters
$\,a, x_0\,$.

The difference between the resonant and nonresonant situation is clearly
visible. In the first case the electron stays in the junction region
and escapes only slowly, while the electron whose momentum is
localized around a slightly different but nonresonant value of
momentum passes the junction ``ballistically''. Wang and Guo$^{22}$
based the mentioned conjecture --- which in a realistic situation
would lead to ultrashort switching times of a few {\em ps} only ---
on a concept of transmitivity of coupled waveguides leaning on a
classical intuition. As we have said in the introduction and
demonstrated in the previous sections, this may be a false guide when
quantum systems are considered. The example of time evolution offers
another illustration. During the resonance--scattering process the
evanescent--mode amplitudes inside the quantum wire are considerably
enhanced; as a result the electron is trapped temporarily inside the
juction. The probability of finding it there in the resonant and
nonresonant case, respectively, is shown on Figure~12. It is
desirable to perform the time--delay analysis for the present model,
in particular, to confirm that the ``switching time'' of the coupler
is inversely proportional to the resonance width.

\section*{Acknowledgement}

The work has been partially supported by the Grants AS No.148409
and GA CR No.202--93-1314.

\vspace{10mm}

\section*{Figure captions}

   \begin{description}
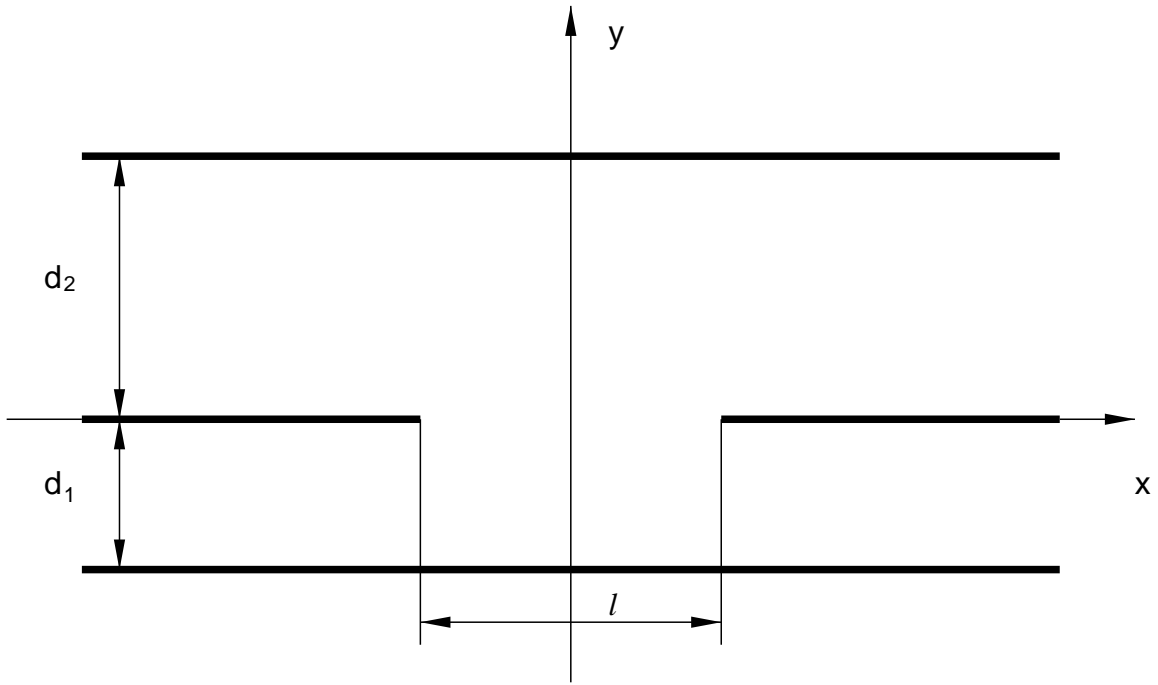
   \item{\bf Figure 1\quad} Laterally coupled quantum waveguides
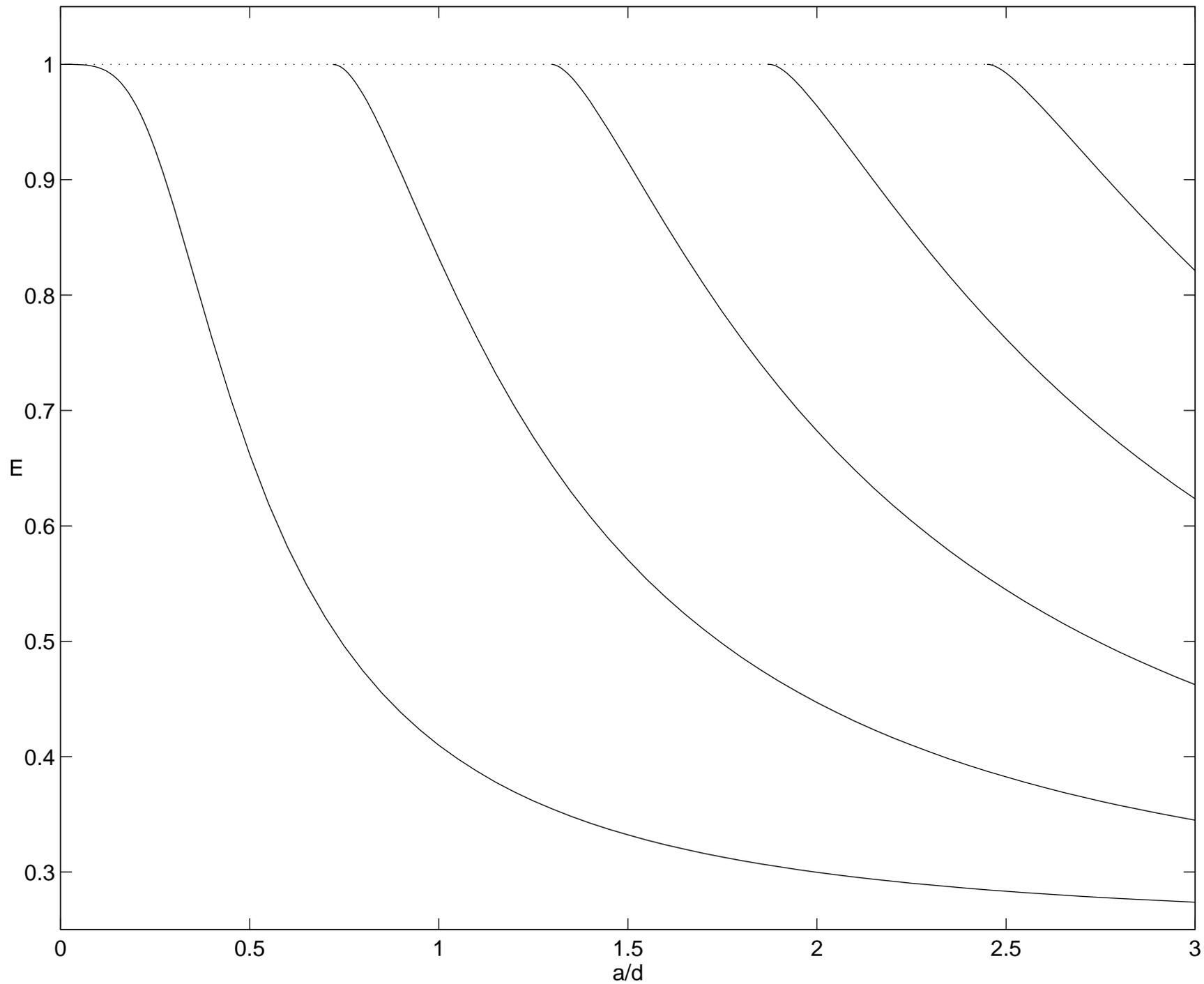
   \item{\bf Figure 2\quad} Bound--state energies {\em vs.} the
window width $\,\ell\,$ in the symmetric case.
   \item{\bf Figure 3\quad} The ground--state eigenfunction in the
symmetric case for $\,\ell/d=0.3\,$.
   \item{\bf Figure 4\quad} The eigenfunction of the second excited
state in the unsymmetric case, $\,\nu=1/2\,$, for $\,\ell/d_1=1.08\,$.
   \item{\bf Figure 5\quad} The conductivity for the particle coming
from the right in the upper duct as a function of the momentum
$\,k\,$ and the width $\,d_2\,$ of the lower tube for $\,d_1=\pi,\;
\ell=2\,$. (a) The particle leaves through the upper left
channel. A deep resonance is clearly visible. (b) The particle leaves
through the lower left channel. The conductivity is zero when there
are no propagating modes in the lower part.
   \item{\bf Figure 6\quad} The quantum probability flow
(\ref{probability flow}) for the
symmetric situation, $\,\nu=1\,$, in the resonance and non--resonance
situation, respectively. The appearance of vortices associated with
the resonance scattering is obvious.
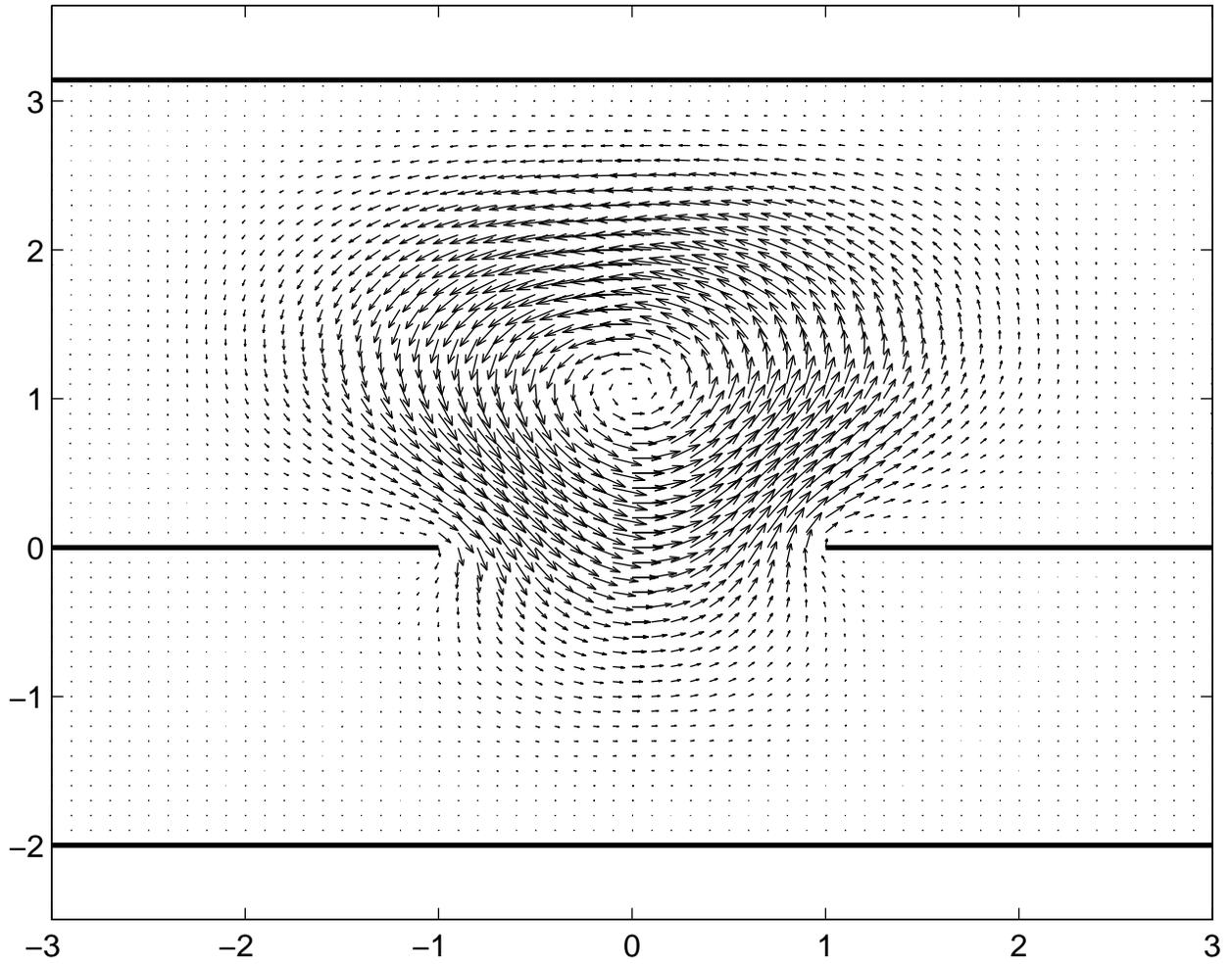
   \item{\bf Figure 7\quad} A single vortex corresponding to the
sharp stopping resonance of Figure~6a. The conductivity is small in
this situation so the waveguide system is closed for the electron
transport.
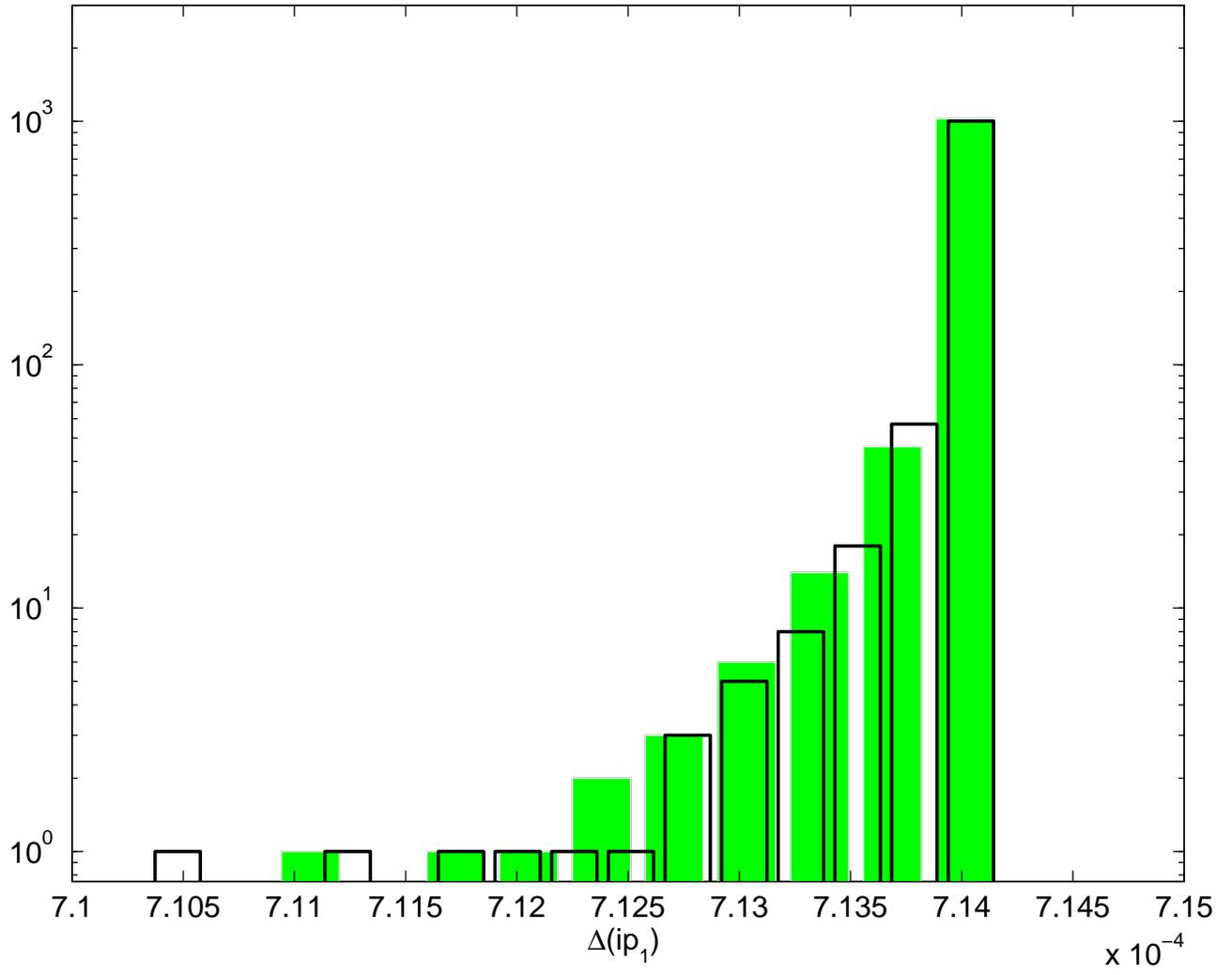
   \item{\bf Figure 8\quad} The unfolded level--spacing distibution
of the symmetric and antisymmetric bound states for $\,\nu=2(1+\sqrt
5)^{-1}\,$.
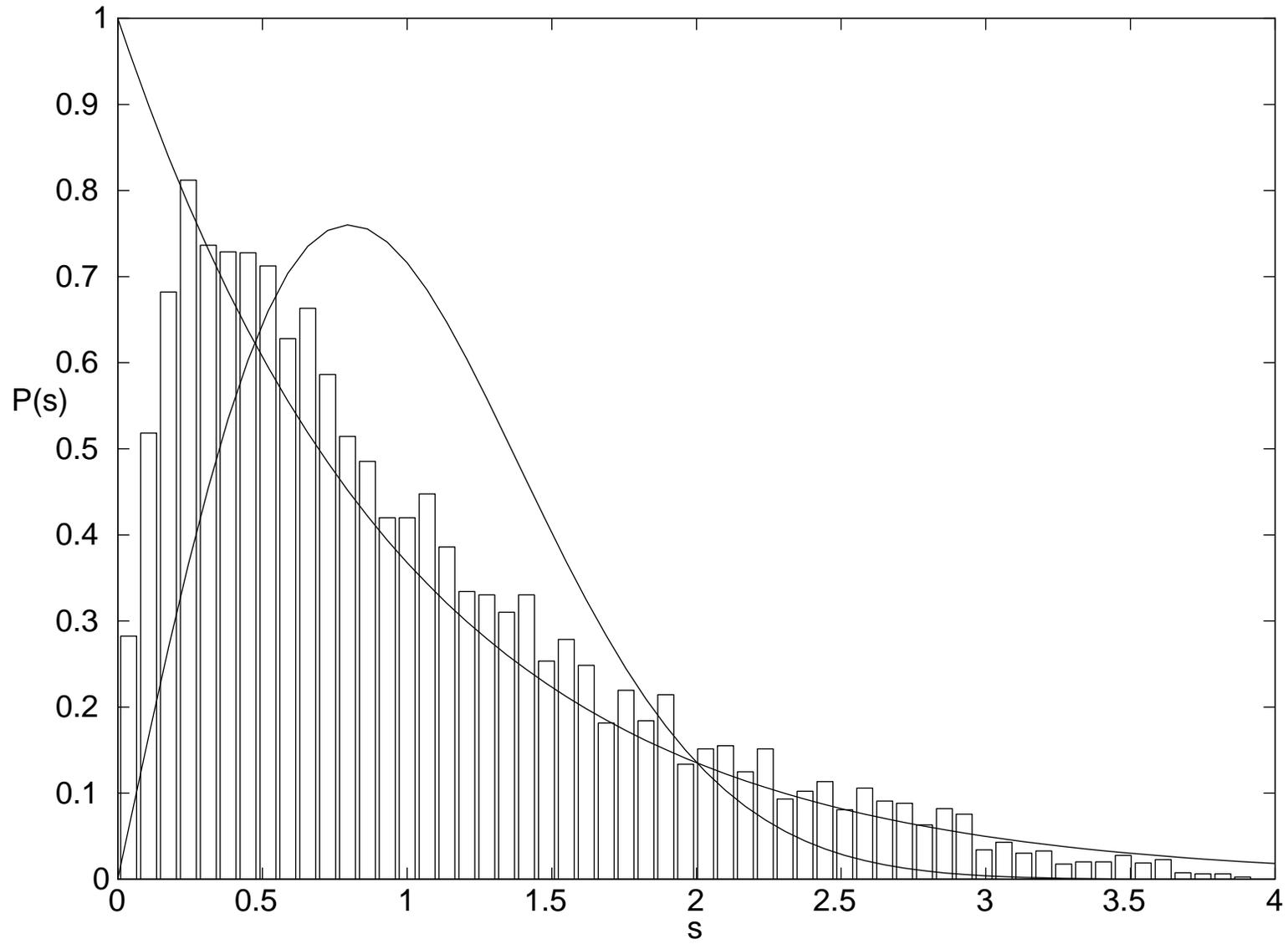
   \item{\bf Figure 9\quad} The unfolded level--spacing distribution
for the S--matrix corresponding to $\,\nu=2/\pi\,$ and averaged over
momentum, in comparison with the Poisson and Wigner distribution.
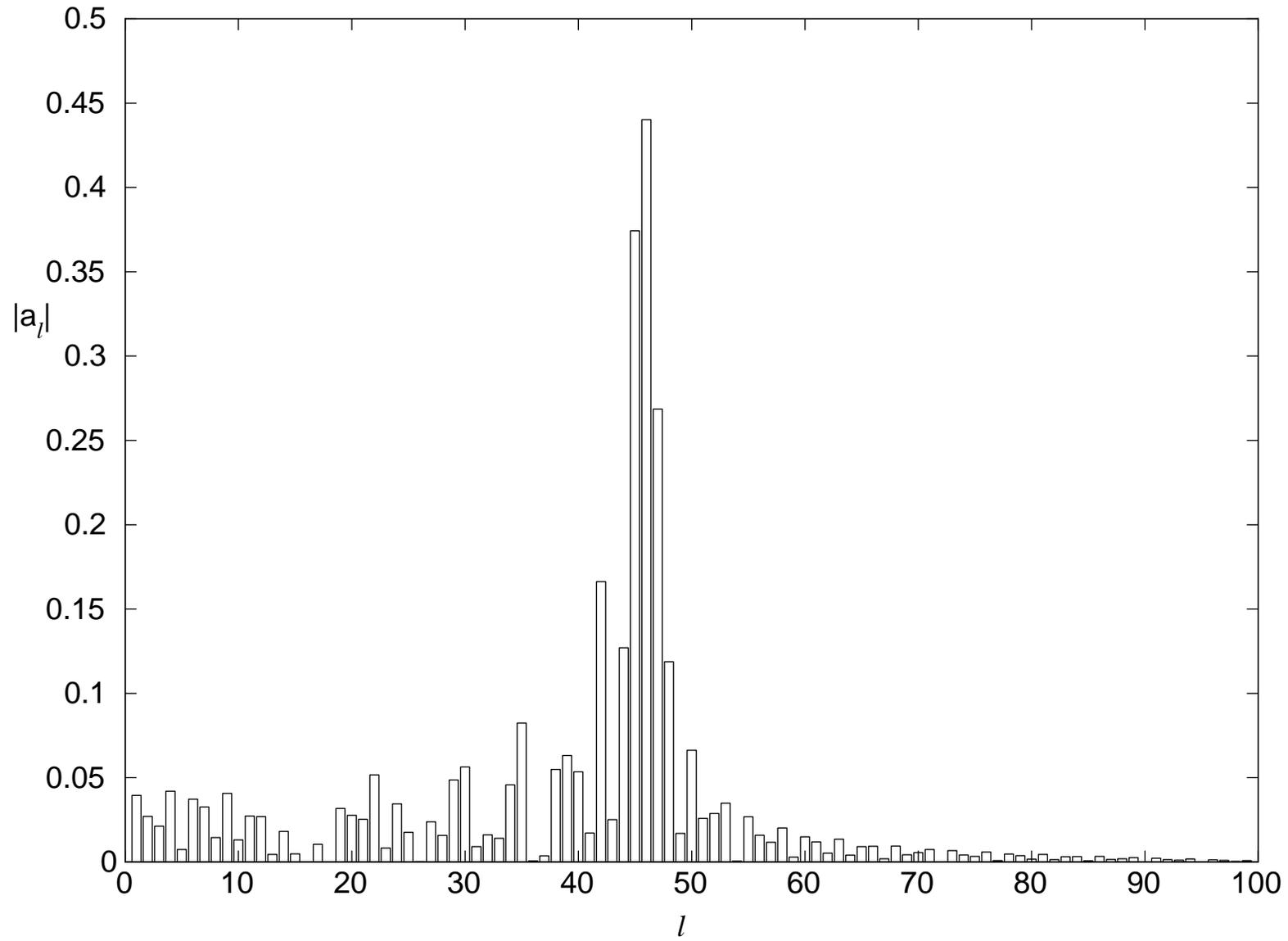
   \item{\bf Figure 10\quad} The absolute value of the coefficients
$\,a_\ell\,$ of eq.(5.2) in the symmetric case for $\,\nu=2/\pi\,$
and $\,k=28.432\;$; the particle is supposed to be initially
in the 18-th transverse mode.
   \item{\bf Figure 11} The time evolution of the wavepacket inside
the junction with $\,\nu=2/\pi\,$ plotted for times
$\,t=0,\,5,\,10,\,15$, and 20, respectively. (a) The resonance case with
$\,k=1.4242\,$, (b) the near--to--resonance situation, $\,k=1.48\,$.
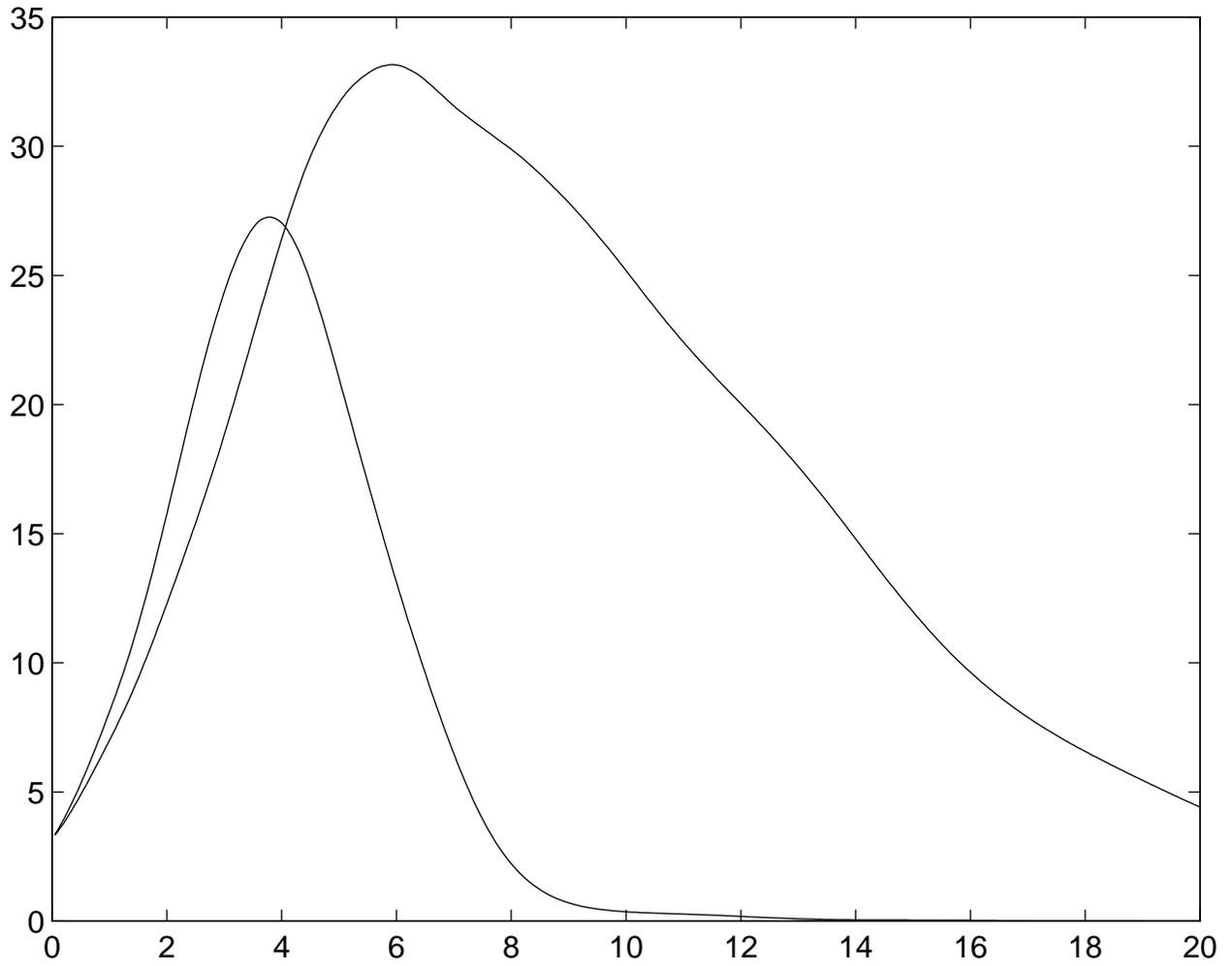
   \item{\bf Figure 12} The probability that the electron will be found
within the junction as a function of time evaluated for the same
parameters as on Figure~11.
   \end{description}

   
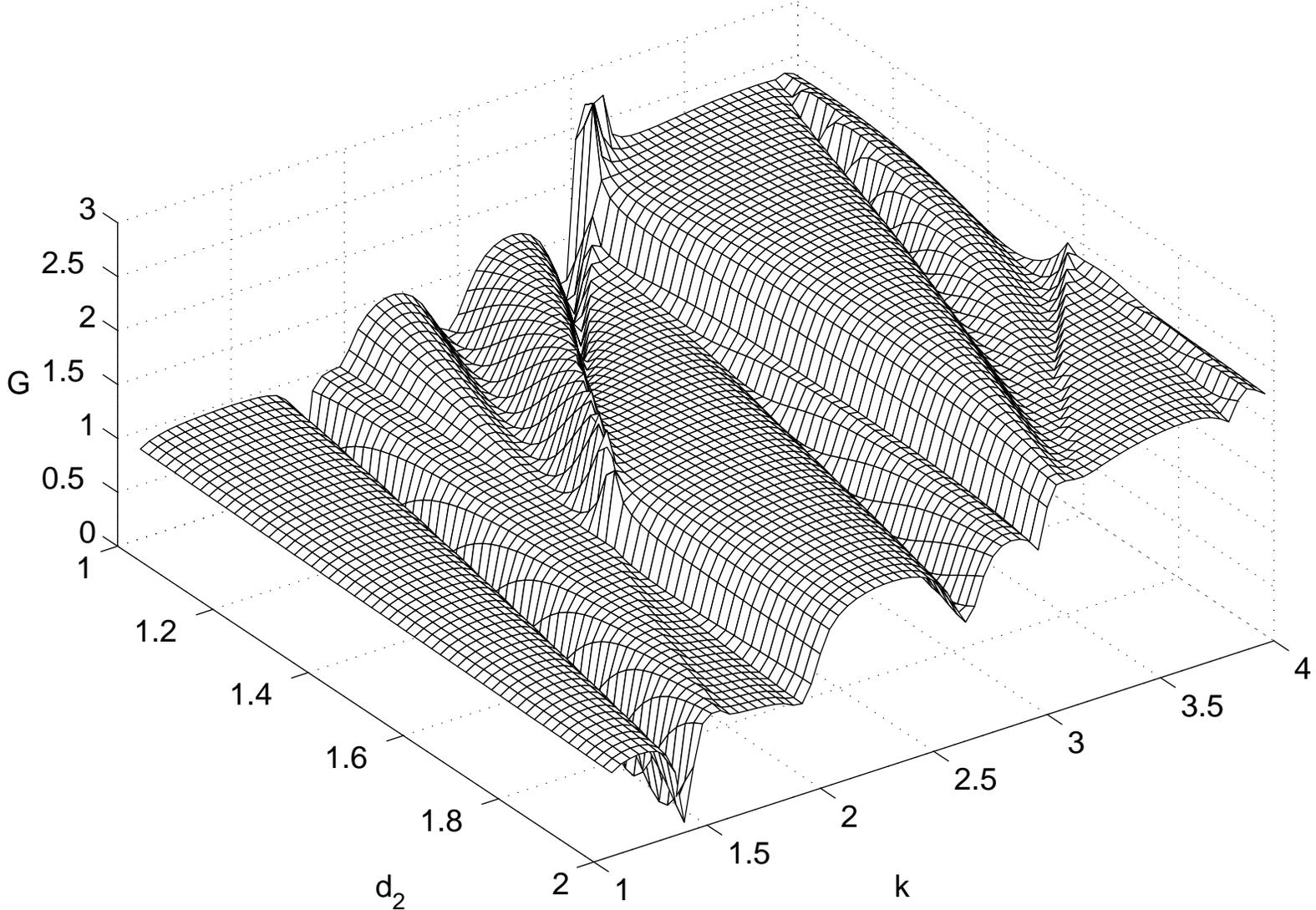
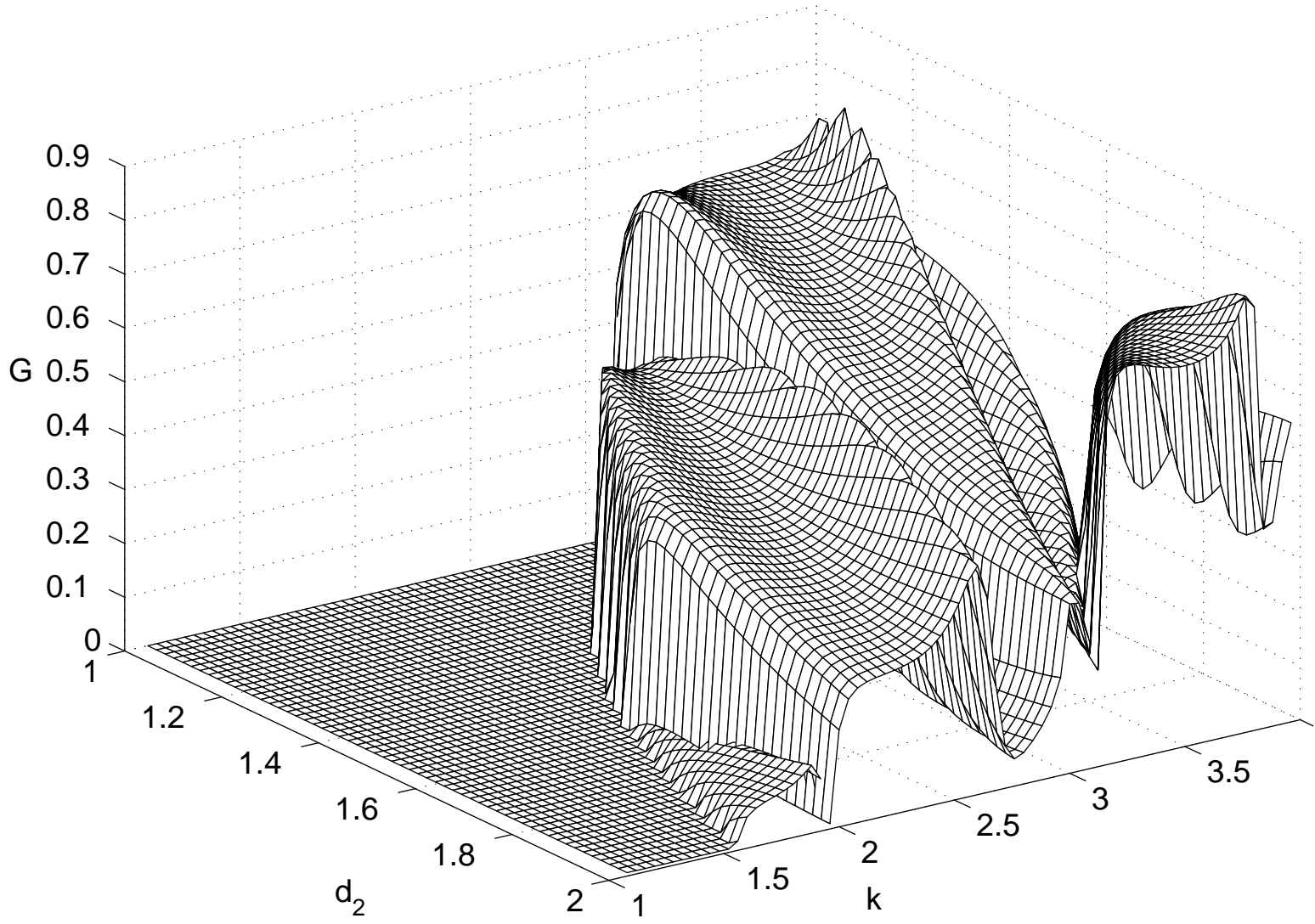
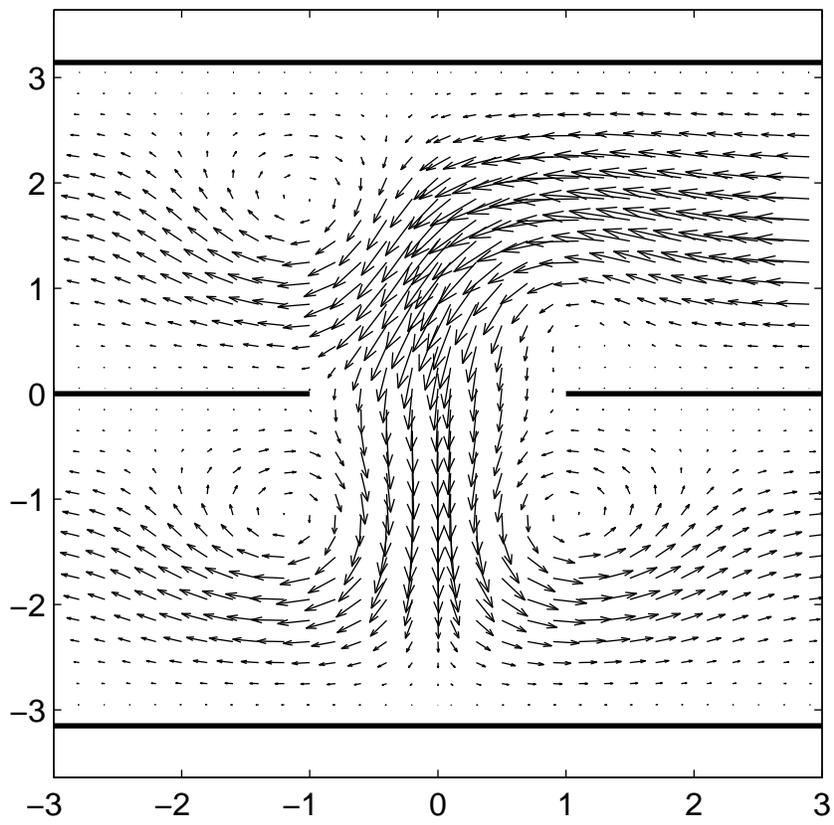
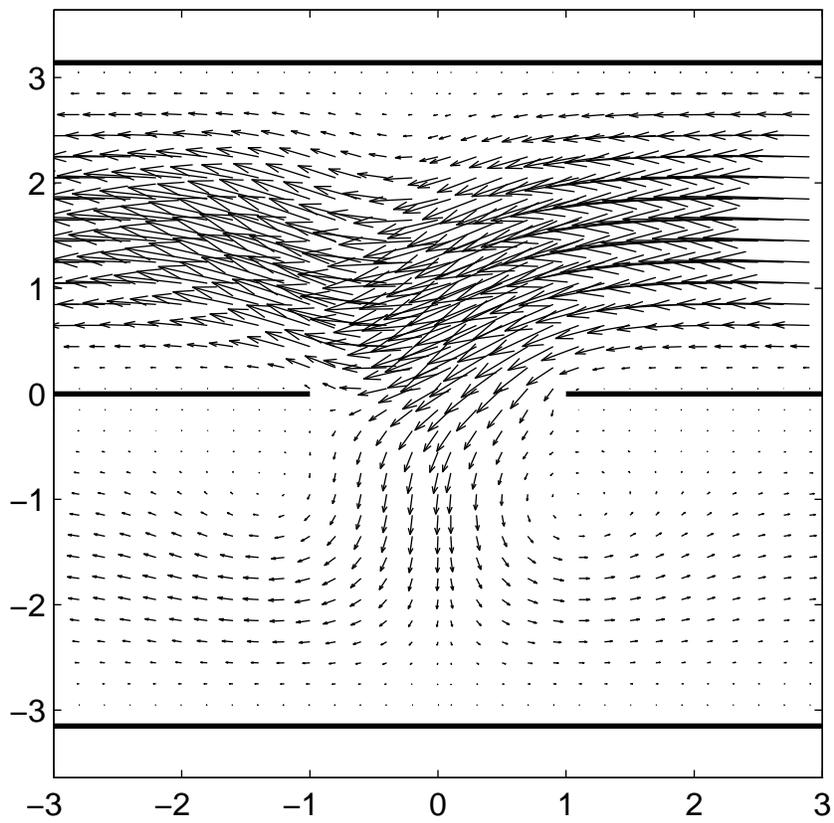
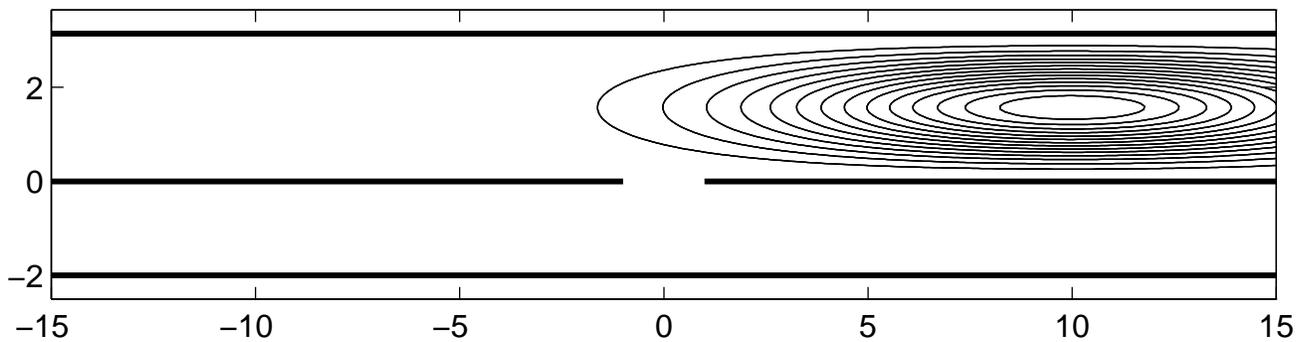
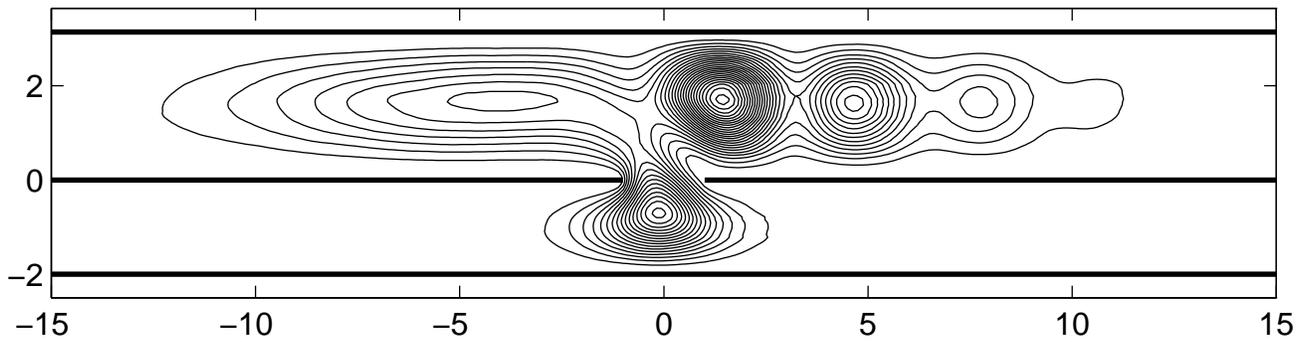
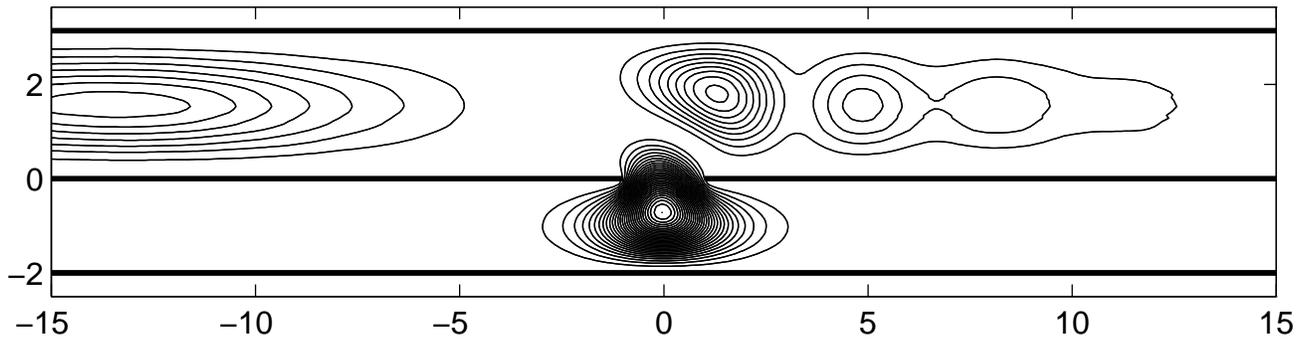
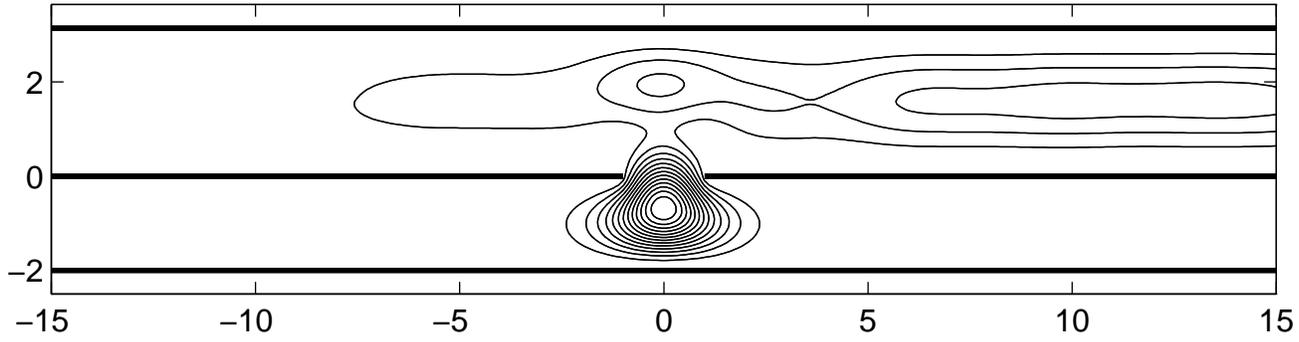
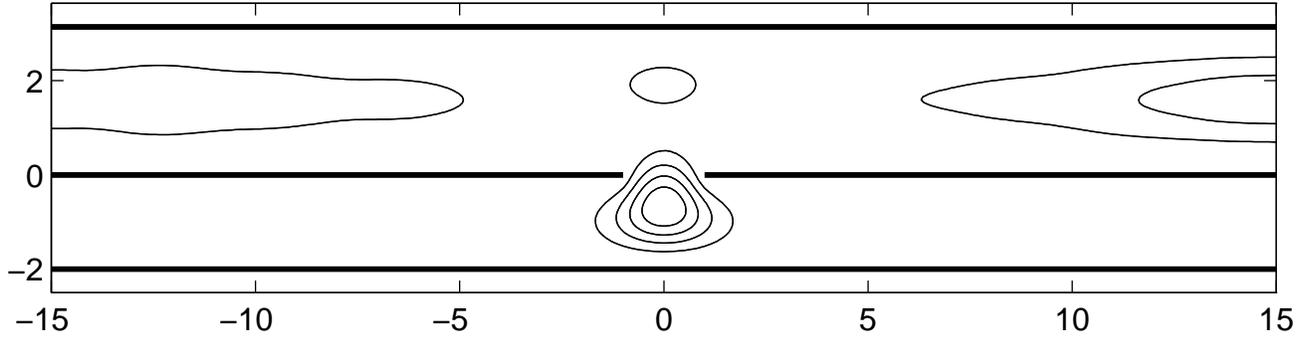
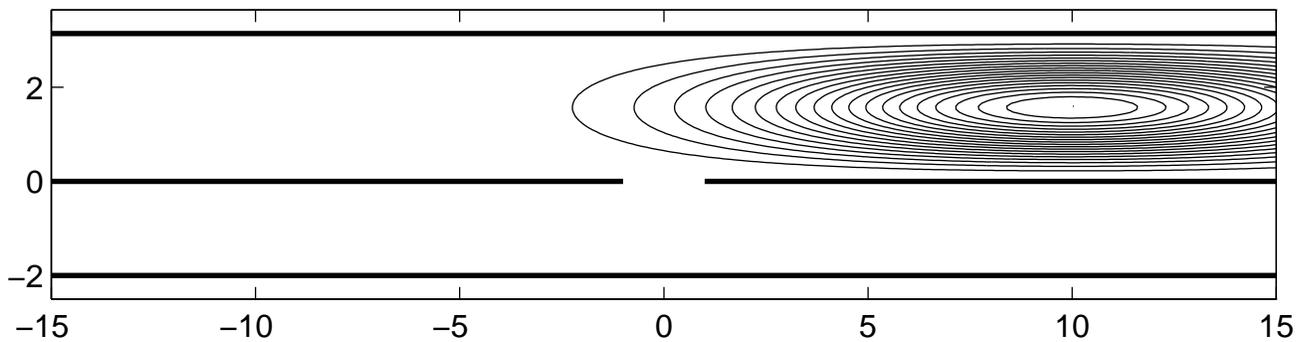
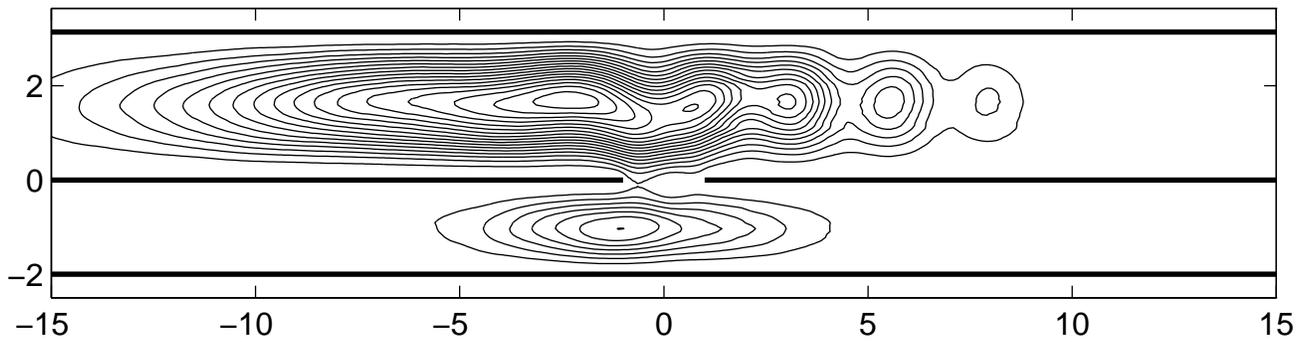
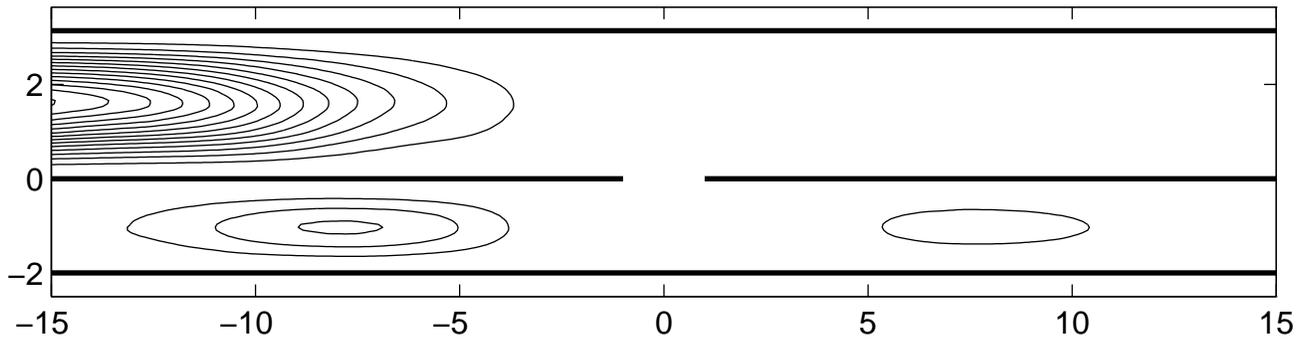
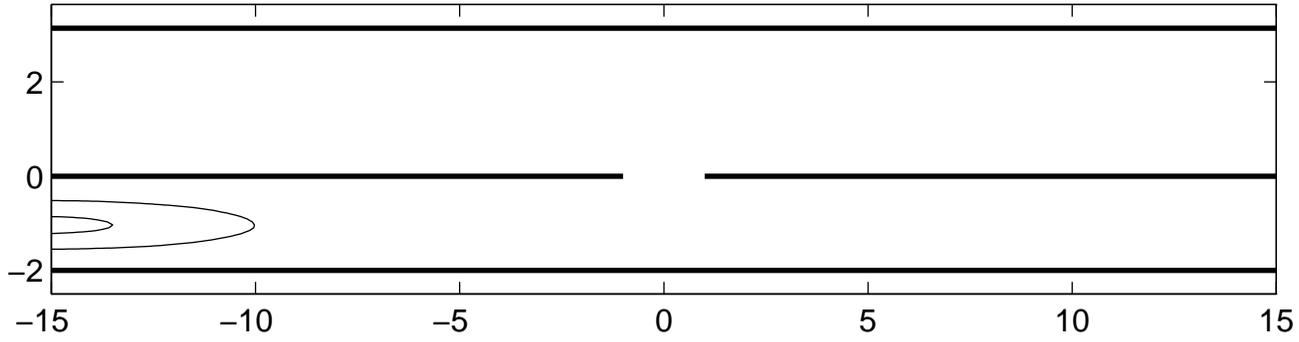
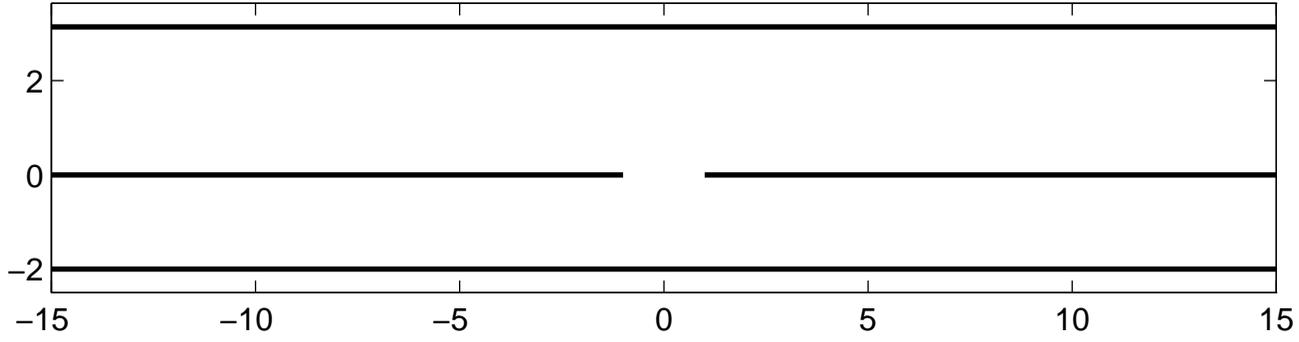
\end{document}